# Alternating Learning Approach for Variational Networks and Undersampling Pattern in Parallel MRI Applications

Marcelo V. W. Zibetti[1], Florian Knoll[2], and Ravinder R. Regatte[1]

[1]Center for Biomedical Imaging, Department of Radiology, New York University Grossman School of Medicine, New York, United States of America

[2] Department of Artificial Intelligence in Biomedical Engineering, Friedrich-Alexander University of Erlangen-Nurnberg, Erlangen, Germany

**ABSTRACT:**

**Purpose:** To propose an alternating learning approach to learn the sampling pattern (SP) and the parameters of variational networks (VN) in accelerated parallel magnetic resonance imaging (MRI).

**Methods:** The approach alternates between improving the SP, using bias-accelerated subset selection, and improving parameters of the VN, using ADAM with monotonicity verification. The algorithm learns an effective pair: an SP that captures fewer k-space samples generating undersampling artifacts that are removed by the VN reconstruction. The proposed approach was tested for stability and convergence, considering different initial SPs. The quality of the VNs and SPs was compared against other approaches, including joint learning methods and VN learning with fixed variable density Poisson-disc SPs, using two different datasets and different acceleration factors (AF).

**Results:** The root mean squared error (RMSE) improvements ranged from 14.9% to 51.2% considering AF from 2 to 20 in the tested brain and knee joint datasets when compared to the other approaches. The proposed approach has shown stable convergence, obtaining similar SPs with the same RMSE under different initial conditions.

**Conclusion:** The proposed approach was stable and learned effective SPs with the corresponding VN parameters that produce images with better quality than other approaches, improving accelerated parallel MRI applications.

**Keywords:** accelerated MRI, image reconstruction, compressed sensing, deep learning, alternating optimization, variational network,

**Word count:** 4530

---

[1] E-mail: Marcelo.WustZibetti@nyulangone.org and Ravinder.Regatte@nyulangone.org
[2] E-mail: florian.knoll@fau.de

# 1. INTRODUCTION

## 1.1 The specific content of this paper:

We propose and validate a new alternated learning approach to jointly learning the sampling pattern (SP) and the parameters of a variational network (VN) [1] used for image reconstruction in accelerated parallel magnetic resonance imaging (MRI) applications. We focus on Cartesian 3D problems used in high-resolution and quantitative MRI, in which the data are collected along multiple k-space lines in a frequency-encoding direction with the SP specifying the 2D phase/partition-encoding positions to be acquired, as illustrated in Figure 1. The acquisition is a 3D process, but we assume that a Fourier transform is applied in the frequency-encoding direction and the volume is separated into multiple slices for 2D reconstructions. Our main research question is whether a joint alternating learning approach for SP and VN can obtain better images than a VN trained with the best fixed SPs recently proposed for MRI, such as a combined variable density and Poisson-disc SPs [2], [3].

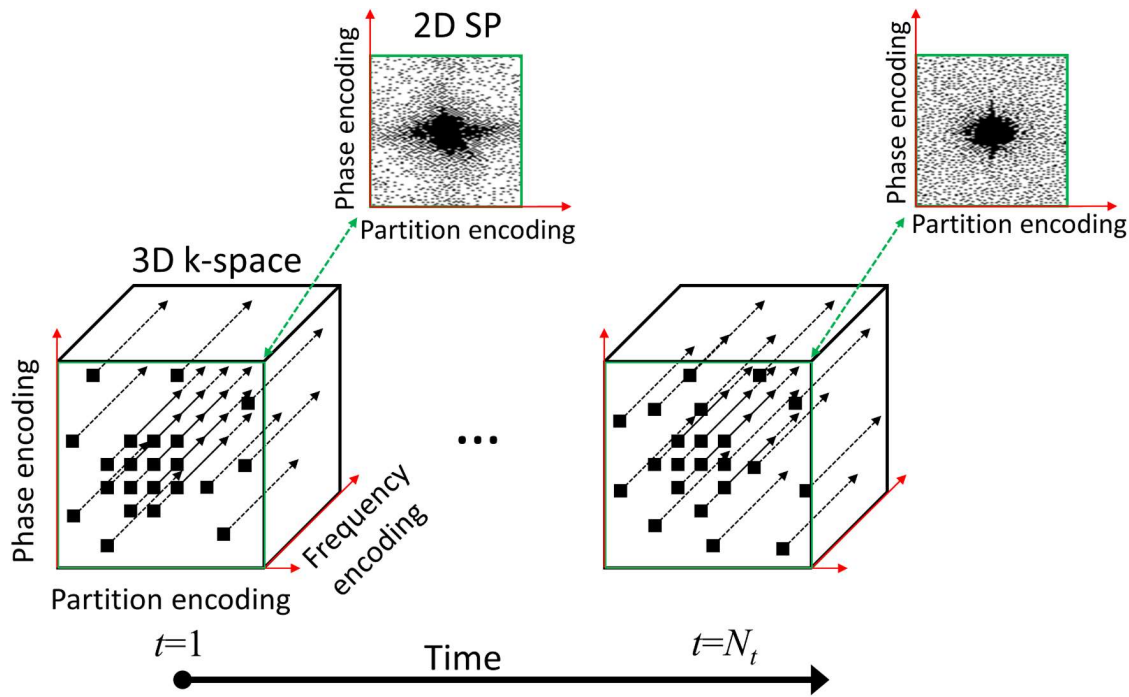

**Figure 1:** Illustration of the 3D+time data collection scheme considered in this work. The SP is in 2D+time, it comprises the time-varying phase and partition encoding positions, for each of which data are to be collected by the MRI scanner for all frequency encoding positions.

## 1.2 Background and purpose:

The acquisition of high-resolution three-dimensional (3D) volumetric data of the human body in MRI is time-consuming. Furthermore, shortening the scan time is necessary for capturing high-resolution images, dynamic processes, quantitative imaging, and reducing health-care costs with increased patient comfort.

The reduction of MRI acquisition time, also known as accelerated MRI, has been obtained by different approaches. Fast magnetic resonance (MR) pulse sequences (FPS) [4]–[6], parallel imaging (PI) [7]–[9], compressed sensing (CS) [10]–[13], and deep learning (DL) image reconstructions [1], [14]–[16] are some examples of the advancements in the accelerated MRI field. These approaches have different mechanisms, FPS can collect more k-space data per unit of time [6], PI uses a receiver array with different coil sensitivities to capture more data in parallel [17] to overcome k-space undersampling, CS relies on incoherent sampling and sparse reconstruction, while DL uses neural networks able to learn non-linear filters to remove artifacts from undersampling.

The quality of image reconstruction in accelerated MRI depends on the sampling process. CS is an example that specific changes in the sampling process [10], [18] are needed to be effective with a specific image reconstruction strategy [19]. Properties such as restricted isometry properties (RIP) and incoherence have guided the design of sampling patterns [10], [12], [20] for sparse reconstructions in MRI. However, new theoretical results [21], [22] showed that incoherence is not a strict requirement in MRI, which are also supported by several studies [23], [24]. In other words, justifying that empirical designs for SP such as variable density [25]–[27] or Poisson-disc [28], [29] (also written as Poisson disk) can also be effective in MRI. Recently, combined variable density and Poisson-disc [2], [3] have been introduced to MRI and it is currently the state-of-art in SP for MRI.

Many accelerated MRI studies have shown that the design of the SP can be customized using reference k-space data of other similar images of particular anatomy [30]–[34]. One class of such approaches, called adaptive sampling approaches, adjusts the probability of the k-space points of variable density SP according to the magnitude of the k-space. Another approach, known as statistical experimental design techniques [35], [36], uses optimization of Cramér-Rao lower bounds for finding good sampling patterns. In both approaches, the particularities of the reconstruction algorithm are not considered in the design of the SP, even though some general formulation is usually assumed.

In data-driven approaches, the customization of the SP is done by optimization algorithms, considering images or datasets containing several images of particular anatomy and a specific method for image reconstruction [37]–[42]. The main idea is that the designed SP should perform well with other images of the same anatomy when the same reconstruction method is used. If one considers that the parameters of the image reconstruction methods can also be learned, then these approaches can be extended to jointly learning the reconstruction and the SP, as seen in [43]–[47]

Most prior data-driven approaches to learning the SP, as in [37]–[40], formulate the SP learning as a subset selection problem and implement it using greedy search algorithms, which lead to extremely high computational costs for SPs of large sizes, particularly the ones used in Cartesian 3D acquisitions. To circumvent this problem, some methods extended deep learning approaches to jointly learn SP and the reconstruction network [43], [44], by implementing the SP as a neural network parameter to be learned. One approach for jointly learning the sampling and reconstruction, called Learning based Optimization of the Under-sampling PattErn (LOUPE), has been originally proposed in [43] for one-coil MRI problems and extended in [48] for multi-coil, or parallel, MRI. Unfortunately, LOUPE does not learn a unique SP, but a sampling density. This way LOUPE does not learn the advantages of the relative position and the distance between k-space samples. For example, studies [2], [3], [28], [29] on the Poisson-disc SPs have shown that the distance between the sampled points is fundamental to the success of SPs used in CS and parallel MRI. In [44], [46], [47], parametric formulation of non-Cartesian k-space trajectories are learned. Although they are very interesting approaches, they require non-uniform Fourier transforms, which is not the best fit for our Cartesian 3D problem described in Section 1.1.

In [42], a fast learning approach for SPs of large sizes was proposed. This approach, called bias-accelerated subset selection (BASS), is still based on the subset selection formulation, as the greedy approaches, but it is as fast as any neural network training algorithm. BASS has shown great results for parallel and CS problems in MRI. However, BASS was never combined with deep learning reconstructions before.

In this sense, we propose an alternated learning approach for the jointly learning SP and reconstruction problem. The SP learning approach is formulated as a subset selection problem solved with BASS, instead of using slow greedy searches. The parameters of the reconstruction network are learned using several short executions of backpropagation algorithms, such as ADAM [49], with post-verification of monotonically decreasing of the cost function. We illustrate that the proposed approach is stable and obtains better SPs, with smaller root mean squared error (RMSE), than other joint learning approaches for the same problem and better than the currently used state-of-art fixed SPs for MRI [2], [3].

## 2. METHODS

### 2.1. Models used:

Parallel MRI methods such as SENSE [50], [51] and many CS approaches [52], including VN, are based on the image-to-k-space forward model, such as:

$$\mathbf{m} = \mathbf{FCx} = \mathbf{Ex}. \tag{1}$$

Here **x** represents 2D+time images, of size $N_y \times N_z \times N_t$, where $N_y$ and $N_z$ are horizontal and vertical dimensions and $N_t$ is the number of time frames, **C** denotes the coil sensitivities transform, which maps **x** into multi-coil-weighted images of size $N_y \times N_z \times N_t \times N_c$, with $N_c$ coils. **F** represents the spatial Fourier transforms (FTs), which are $N_t \times N_c$ repetitions of the 2D-FT, and **m** is the fully sampled data, of size $NN_s$, where $N = N_y N_z N_t$ and $N_s = N_c$. Both systems combine in the encoding matrix **E**. When accelerated MRI by undersampling is used, then the SP is included in the model as:

$$\bar{\mathbf{m}} = \mathbf{S}_\Omega \mathbf{FCx} = \mathbf{E}_\Omega \mathbf{x}, \tag{2}$$

where $\mathbf{S}_\Omega$ is the sampling function using the SP specified by $\Omega \subset \Gamma$ (same for all coils) which is a subset of the fully sampled Cartesian set $\Gamma$, and $\bar{\mathbf{m}}$ is the undersampled multi-coil k-space data (or k-t-space when $N_t > 1$), with $MN_s$ elements, where $M$ is the number of sampled points in the undersampled k-space. The acceleration factor (AF) in MRI is given by $N/M$. Depending on the approach used to obtain coil sensitivity, additional sampling might be necessary. Here, we assume a central area of the k-space will be always fully sampled and used to compute coil sensitivities with auto-calibration methods, such as [53].

### 2.2. Variational networks for image reconstruction:

The VNs are inspired by the minimization problem [54], given by

$$\hat{\mathbf{x}} \in \arg\min_{\mathbf{x}} \|\bar{\mathbf{m}} - \mathbf{E}_\Omega \mathbf{x}\|_2^2 + \sum_{f=1}^{N_f} \langle \Phi(\mathbf{K}_f \mathbf{x}), \mathbf{1} \rangle, \tag{3}$$

but instead, it approximates a solution by a fixed number of iterations, denoted by $J$, of a gradient descent-like algorithm [1], given by

$$\mathbf{x}_{j+1} = \mathbf{x}_j - \left( \alpha_j \mathbf{E}_\Omega^* (\bar{\mathbf{m}} - \mathbf{E}_\Omega \mathbf{x}_j) + \sum_{f=1}^{N_f} \mathbf{K}_{j,f}^b \Phi'(\mathbf{K}_{j,f} \mathbf{x}_j) \right), \tag{4}$$

where the complex-valued vector **x** and the encoding matrix $\mathbf{E}_\Omega$ are described in (1) and (2). Also, $1 \leq j \leq J + 1$ represents the iteration index, where $J = 10$ was chosen in this work. In the context of deep learning [55], [56], it corresponds to the number of layers.

All the VN parameters, i.e. convolutional filters $\mathbf{K}_{j,f}^b, \mathbf{K}_{j,f}$ ($N_f = 24$) and step-sizes $\alpha_j$, are learned from data [1]. In this study, activation functions $\Phi'$ are fixed rectified linear units (ReLu) [55]. The set of convolutional filters $\mathbf{K}_{j,f}^b, \mathbf{K}_{j,f}$ are spatio-temporal filters of size $11 \times 11 \times N_t$, which means convolutional filters correlate all temporal frames.

Note that the VN in equation (4) resembles a general iterative regularized reconstruction algorithm. The central term in (4), $\alpha_j \mathbf{E}_\Omega^* (\bar{\mathbf{m}} - \mathbf{E}_\Omega \mathbf{x}_j)$, is responsible for reducing k-space error, while the right term in (4), $\sum_{f=1}^{N_f} \mathbf{K}_{j,f}^b \Phi'(\mathbf{K}_{j,f} \mathbf{x}_j)$, reduces undesired artifacts in the image. The spatio-temporal filters used by the VN are implemented as convolutional filters learned from training data. These components are different for each layer $j$. The output of a VN can be written as:

$$\hat{\mathbf{x}} = R_\theta(\bar{\mathbf{m}}, \Omega), \tag{5}$$

where $R_\theta$ represents the VN reconstruction with network parameters $\theta = \left\{\{\mathbf{K}^b_{j,f}, \mathbf{K}_{j,f}\}_{f=1}^{N_f}, \alpha_j\right\}_{j=1}^{J}$, input sampling pattern $\Omega$ and undersampled data $\bar{\mathbf{m}} = \mathbf{S}_\Omega \mathbf{m}$, and output reconstructed image $\hat{\mathbf{x}}$. Note that coil sensitivity maps used in $\mathbf{C}$ (that is part of $\mathbf{E}_\Omega$) are also input parameters for image reconstruction, but since they are obtained from $\bar{\mathbf{m}}$, and they are not considered a learned parameter, we will not make them explicit in (5).

### 2.3. Joint learning the sampling pattern and image reconstruction:

The joint learning process can be formulated through the following criterion:

$$\hat{\Omega}, \hat{\theta} = \underset{\substack{\Omega \subset \Gamma, \theta \in \Theta \\ s.t. |\Omega|=M}}{\operatorname{argmin}} \frac{1}{N_i} \sum_{i=1}^{N_i} f(\mathbf{x}_i, R_\theta(\bar{\mathbf{m}}_i, \Omega)), \qquad (6)$$

where $R_\theta(\bar{\mathbf{m}}_i, \Omega) = \hat{\mathbf{x}}_i$ is the reconstructed image using undersamped data $\bar{\mathbf{m}}_i$, with sampling pattern $\Omega$ and network parameters $\theta$. Note that $i$ represent the index of the image and k-space data in the dataset with $N_i$ elements. Note that undersampled data can be synthetically generated using $\bar{\mathbf{m}}_i = \mathbf{E}_\Omega \mathbf{x}_i$, if images and coil sensitivities are available, or reference images can be produced by $\mathbf{x}_i = R(\mathbf{m}_i, \Gamma)$, as a reference for fully sampled image reconstruction (which do not necessarily need to be a deep learning reconstruction).

The criterion (6) is defined in the image domain. This approach is useful when fully-sampled k-space data are not directly available, but images reconstructed from fully-sampled data and coil sensitivities are available, or when one wants to weigh differently particular regions of interest of the image to obtain better quality in more relevant structures. In this study, we will consider $f(\mathbf{x}_i, R_\theta(\bar{\mathbf{m}}_i, \Omega)) = \|\mathbf{x}_i - R_\theta(\mathbf{E}_\Omega \mathbf{x}_i, \Omega)\|_2^2$, since it is a criterion easily available in most deep learning tools and it is the same criterion used in the other compared approaches.

## 3. PROPOSED ALTERNATING LEARNING APPROACH:

While end-to-end joint learning is appealing, the approximations required to adapt a non-differential problem, such as SP learning, in a backpropagation algorithm, as done by LOUPE, may lead to instabilities in the learning algorithm. To obtain more stability regarding the choice of the SP and less dependence on the initial guess of the SP and network parameters, we propose an alternated approach for this joint learning process.

The proposed alternated learning approach uses a combination of short executions of a subset selection solver for the SP and short execution of a backpropagation-learning algorithm for the neural network parameters. Both use cost function checks after execution to ensure a monotonical decrease of the cost function. For fast solving the subset selection problem, the BASS algorithm [42] was chosen, and for neural network parameters learning the ADAM algorithm was chosen [49]. An overview of the proposed approach is shown in Figure 2. BASS has been proven effective in learning the SP for compressed sensing and parallel MRI methods before, however it was never tested with deep learning algorithms. ADAM is currently one of the most used approaches to train deep neural networks.

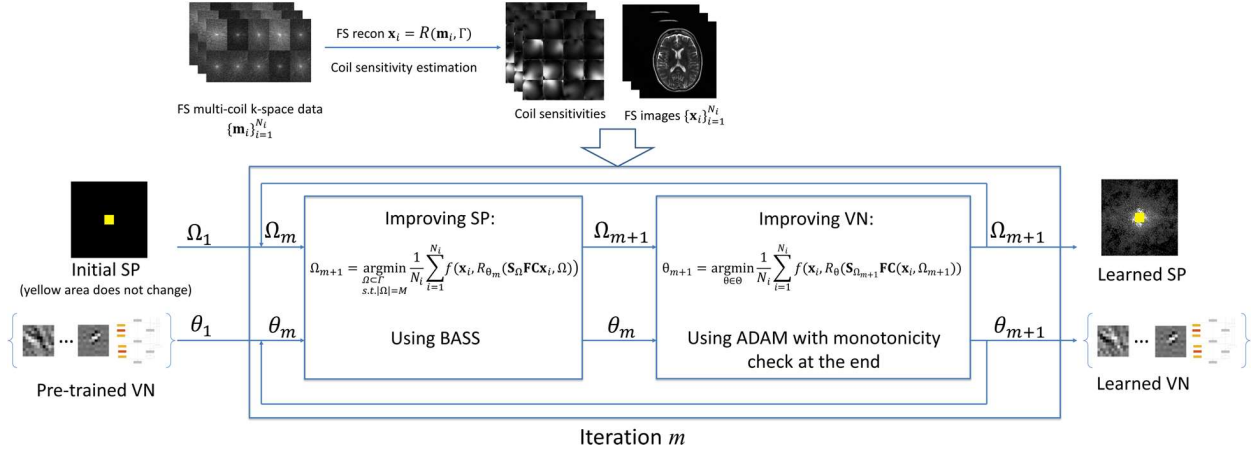

**Figure 2:** General overview of the proposed approach for jointly learning the SP and the parameters of a VN on Parallel MRI problems.

### 3.1. Formulation:

The proposed approach to perform (6) is:

$$\Omega_{m+1} = \underset{\substack{\Omega \subset \Gamma \\ s.t. |\Omega|=M}}{\operatorname{argmin}} \frac{1}{N_i} \sum_{i=1}^{N_i} f(\mathbf{x}_i, R_{\theta_m}(\mathbf{S}_\Omega \mathbf{FC}\mathbf{x}_i, \Omega)) , \qquad (7)$$

$$\theta_{m+1} = \underset{\theta \in \Theta}{\operatorname{argmin}} \frac{1}{N_i} \sum_{i=1}^{N_i} f(\mathbf{x}_i, R_\theta(\mathbf{S}_{\Omega_{m+1}} \mathbf{FC}(\mathbf{x}_i, \Omega_{m+1})) , \qquad (8)$$

Where some iterations of BASS are used to solve (7), and some iterations of ADAM are used to solve (8).

#### 3.1.1. Forced monotonicity of the joint learning approach:

In order to improve the stability of the joint learning approach, monotonicity is forced in the alternated procedure. BASS, used to solve Equation (7), is originally proposed as a monotone algorithm [42], as shown in the next section. However, ADAM that is used for neural network learning by solving Equation (8), is not monotone. This means that the parameters learned in a particular iteration could be worse than the previous $\theta_m$ (this strongly depends on the parameters used in ADAM, including the number of iterations). In this case, after a short execution of ADAM, the previous parameters are maintained in case the new learned parameters do not reduce the cost function.

### 3.2. Bias-Accelerated Subset Selection:

BASS is a heuristic optimization approach for subset selection based on bit change, such as Pareto optimization for subset selection (POSS) [57]–[59]. In general, bit change algorithms are much faster than greedy algorithms. While greedy algorithms compute many candidates before choosing the one with maximum improvement, bit-wise mutation algorithms change the current best SP, adding and removing some elements of it, according to some heuristic rule, accepting the new solution if it is better. BASS is faster because it uses heuristics rules to choose the new k-space points based on large observed errors in the k-space [42].

The learning process can start with any known SP, such as Poisson-disc, variable density, or even an empty SP. It is also possible to constraint some regions of the SP, such as the central area that is used for coil sensitivity auto-calibration. BASS, originally proposed in [42], is detailed in Algorithm 1, for completeness, where the cost function $F(\Omega) = \frac{1}{N_i}\sum_{i=1}^{N_i}\|\mathbf{e}_i\|_2^2$, for $\mathbf{e}_i = \mathbf{x}_i - R_\theta(\mathbf{E}_\Omega \mathbf{x}_i, \Omega)$, is different from [42] in this study. When monotonicity is disabled, then line 10 of Algorithm 1 is replaced by $\Omega \leftarrow \Omega'$.

---

**Algorithm 1**: BASS $O(\Omega_{init}, K, \alpha, L)$,

1. $\Omega \leftarrow \Omega_{init}$
2. $K \leftarrow K_{init}$
3. $l \leftarrow 1$
4. **While** $l \leq L$ **do**
5.     $\Omega_r \leftarrow$ **select-remove**$(\Omega, K, \rho_r(K, M, l))$
6.     $\Omega_a \leftarrow$ **select-add**$(\Omega, K, \rho_a(K, M, N, l))$
7.     $\Omega' \leftarrow (\Omega \setminus \Omega_r) \cup \Omega_a$
8.     **if** $|\Omega| \neq M, \Omega \leftarrow \Omega'$
9.     **if** $|\Omega| = M$
10.       **if** $F(\Omega') \leq F(\Omega), \Omega \leftarrow \Omega'$
11.       **else** $K \leftarrow \lfloor (K-1)\alpha \rfloor + 1$
12. $l \leftarrow l + 1$
13. **return** $\Omega$

---

#### 3.2.1. Modification in the heuristics used in BASS:

In BASS, the elements of $\Omega_a$ and $\Omega_r$ are selected by the functions **select-add** and **select-remove** in similar ways. The functions **select-add**$(\Omega, K, \rho_a(K, M, N))$ and **select-remove**$(\Omega, K, \rho_r(K, M))$ in Algorithm 1 are used to modify $\Omega$ in the composition of a new candidate $\Omega'$. The number $K$ of elements to be added/removed varies with iteration. For $1 \leq i \leq N_i$, let $\mathbf{e}_i = \mathbf{FC}(\mathbf{x}_i - R_{\theta_m}(\mathbf{E}_\Omega \mathbf{x}_i, \Omega))$. Note that each of the $N$ components of $\mathbf{e}_i$ is an $N_s$-dimensional vector. For $1 \leq k \leq N$, and $1 \leq s \leq N_s$, where $[\mathbf{e}_i]_{k,s}$ denotes the $s$-th component (corresponding to the coil) of the $k$-th component (corresponding to the k-space position) of $\mathbf{e}_i$. For **select-add**, we define the measure of importance (MI) used in this work, for $1 \leq k \leq N$, as

$$\boldsymbol{\varepsilon}_k = \frac{1}{N_i N_s} \sum_{i=1}^{N_i} \sum_{s=1}^{N_s} |[\mathbf{e}_i]_{k,s}|^2, \tag{9}$$

which is referred to as the **ε**-map. The purpose of **select-add** is to select $K$ elements from $\Gamma \backslash \Omega$ in a randomly biased manner [42]. For **select-remove**, a sequence with $K$ elements that are in $\Omega$, is generated in the same way, but using $\mathbf{r}_k$ as the MI, instead of $\boldsymbol{\varepsilon}_k$. The MI for **select-remove** is

$$\mathbf{r}_k = \frac{1}{N_i N_s} \sum_{i=1}^{N_i} \frac{\sum_{s=1}^{N_s} |[\mathbf{e}_i]_{k,s}|^2 + \delta}{\sum_{s=1}^{N_s} |[\mathbf{m}_i]_{k,s}|^2 + \delta}, \tag{10}$$

for $1 \leq k \leq N$, which is referred to as the **r**-map, with $\delta$ a small constant to avoid zero/infinity in the definition of $\mathbf{r}_k$.

The probability of pre-selecting elements should respect pre-defined ranges, $K/M < \rho_r(K, M, l) \leq 1$ and $K/(N - M) < \rho_a(K, M, N, l)) \leq 1$. We used constant probabilities for $\rho_a = 1/4$ and $\rho_r = 1/4$, and we used the same positional constraints (PC) rules proposed in [42]. The most expensive part of **select-add** and **select-remove** is the computation of the recoveries given by $R_{\theta_m}(\mathbf{E}_\Omega \mathbf{x}_i, \Omega)$. However, these computations are cheap because they are just forward propagations of a VN, which are much faster than iterative CS reconstructions [60].

## 4. EXPERIMENTS:

### 4.1. Datasets:

In these experiments, we utilized two MRI datasets to test the proposed approach. One dataset, denominated "brain", contains 320 brain T$_2$-weighted images from the fast MRI dataset [61]. Of these, 260 were used for training and 60 for testing. The k-space data has a size $N_y \times N_z \times N_t \times N_c = 320 \times 320 \times 1 \times 16$, and the reconstructed images are $N_y \times N_z \times N_t = 320 \times 320 \times 1$. The second dataset, denominated "knee joint", contains 320 T$_{1\rho}$-weighted knee images for quantitative T$_{1\rho}$ mapping, of size $N_y \times N_z \times N_t \times N_c = 128 \times 64 \times 10 \times 16$, and the reconstructed images are $N_y \times N_z \times N_t = 128 \times 64 \times 10$. Of these, 260 were used for training and 60 for testing. The k-space data for all images are normalized by the largest component.

### 4.2. Pre-training the VNs:

To avoid random initializations of the VN parameters affecting the final results, the same pre-trained VN parameters were used in all cases. For pre-training, randomly generated SPs of all kinds, including Poisson-disc, variable density, uniform density, and combined variable density with Poisson-disc (VD+PD) were used. For pre-training, we used 80 epochs of the ADAM algorithm [49] with an initial learning rate of $2\times10^{-4}$, with a learning rate drop factor of 0.5 applied every 5 epochs, and a batch size of 8 images.

### 4.3. Sampling patterns and learned VN used for comparison:

The proposed alternated approach was compared with the following approaches:
- **Pre-trained VN, VD+PD SP:** the pre-trained VN was tested with a combined variable density and Poisson-disc (VD+PD) [2] SP.
- **Re-trained VN, VD+PD SP:** the VN was re-trained (with the pre-trained VN as initialization) with one fixed VD+PD SP, and it was tested with the same VD+PD SP.
- **VN trained with LOUPE, LOUPE SP:** the VN was re-trained with LOUPE approach [43], [48], where the sampling density is learned, and it was tested with one realization of the SP from the sampling density, named LOUPE SP.
- **Re-trained VN, LOUPE SP:** the VN was re-trained (with the VN trained with LOUPE as initialization) but using one fixed LOUPE SP, and it was tested with the same LOUPE SP.

### 4.4. Configuration of the training process:

The proposed alternating learning algorithm consists of alternating executions of BASS and ADAM. Each run of BASS starts with $K_{init}=1024$ and $\alpha=0.5$ and stops when $K=1$. Each run of ADAM uses 8 epochs with an initial learning rate of $2\times10^{-4}$, with a learning rate drop factor of 0.25 applied every 2 epochs, and batch size of 8 images. The alternating procedure is repeated until the cost function stops decreasing for 5 cycles or a maximum of 1000 cycles passed. For LOUPE, we used 80 epochs ADAM with an initial learning rate of $2\times10^{-4}$, with a learning rate drop factor of 0.5 applied every 5 epochs, and a batch size of 8 images. For re-training a VN with a fixed SP, such as a VD+PD SP or LOUPE SP, we used 80 epochs ADAM with an initial learning rate of $2\times10^{-4}$, with a learning rate drop factor of 0.5 applied every 5 epochs, and a batch size of 8 images.

### 4.5. Assessment:

The quality of the results obtained with the SP was evaluated using the root mean squared error (RMSE), defined as:

$$RMSE\left(\{\mathbf{x}_i\}_{i=1}^{N_v}, \{\hat{\mathbf{x}}_i\}_{i=1}^{N_v}\right) = \sqrt{\frac{1}{N_t N_i} \sum_{i=1}^{N_v} \|\mathbf{x}_i - \hat{\mathbf{x}}_i\|_2^2}. \qquad (11)$$

## 5. RESULTS:

### 5.1. Convergence and stability of the proposed approach:

We tested the stability of the proposed approach in terms of convergence with forced monotonicity and with different initial SP. These two tests illustrate the convergence and the stability of the proposed approach.

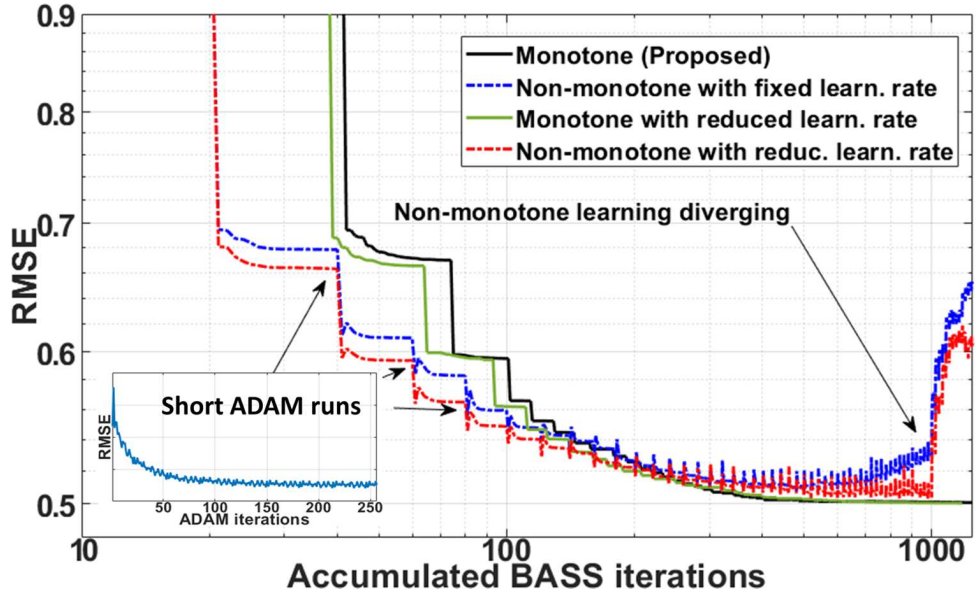

**Figure 3:** Convergence of the proposed approach with monotonicity enabled (proposed), disabled (Non-monotone), and when reduced initial learning rate of the short ADAM runs (factor 0.95, applied on every new ADAM run) is used. In this figure, the discontinuities are due to short runs of ADAM algorithm, only three runs are marked with arrows and an illustrative convergence curve for ADAM is shown in the small box.

In Figure 3 we illustrate the RMSE over the accumulated BASS iterations (in logarithmic scale) of the proposed approach when forced monotonicity is enabled and disabled (for both: BASS and ADAM). We also compared both methods with a fixed or reducing initial learning rate of ADAM, as a way to avoid divergence. Note that initial convergence is faster when monotonicity is disabled than when it is enabled, but later on, in the iterations, the RMSE increases again if no monotonicity is forced. Reducing the initial learning rate does not prevent this divergence.

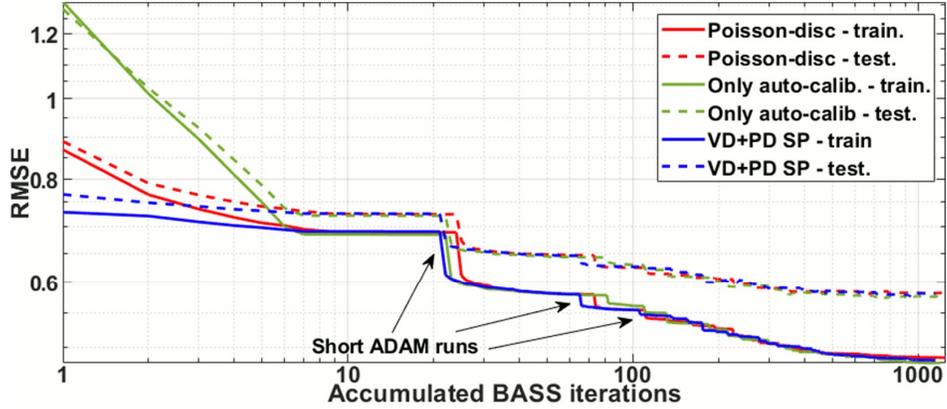
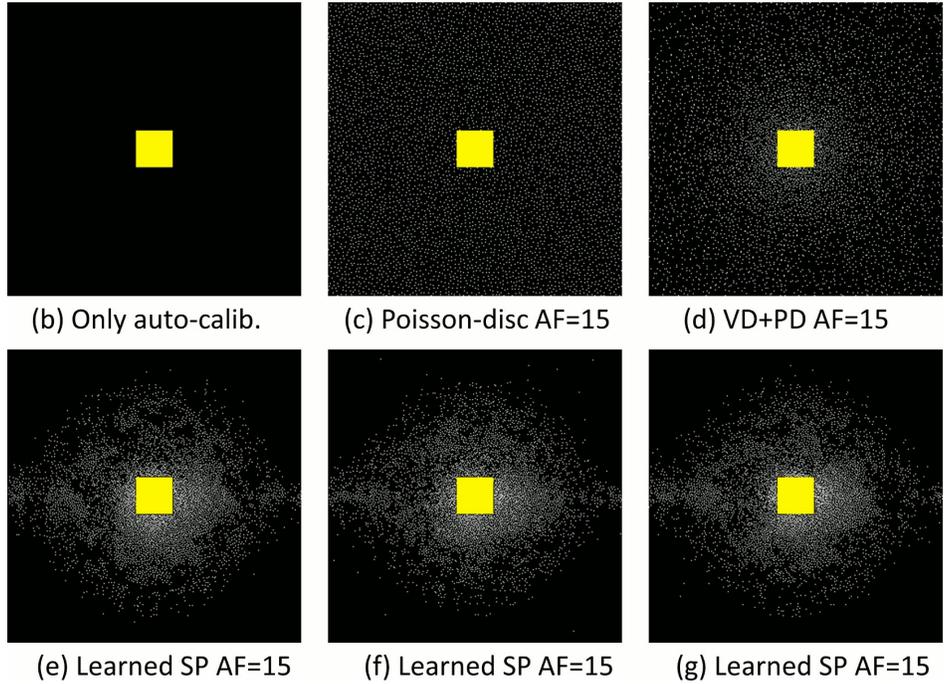

**Figure 4:** Convergence of the proposed approach with different initial SPs, but with the same pre-trained VN. In (a), the RMSE over accumulated BASS iterations are shown (in logarithmic scale). In (a), the discontinuities are due to short runs of ADAM algorithm. The initial SPs are shown in (b) an empty SP, only with the fixed auto-calibration area in yellow, in (c) a Poisson-disc SP, and in (d) a VD+PD SP. The corresponding learned SPs are shown in (e)-(g).

In Figure 4 we illustrate the learned SPs considering different initial SPs: an empty SP, only with the fixed central auto-calibration area, shown in Figure 4(b), a Poisson-disc SP, shown in Figure 4(c), and the VD+PD SP, in Figure 4(d). The RMSE curves for the training and testing datasets are shown in Figure 4(a), while the resulting learned SPs with AF=15 are shown in figures 4(e)-(g), respectively. All resulting learned sets, SP and VN, obtained a final RMSE of 0.58 for the testing dataset.

## 5.2. Comparison against the tested approaches:

We compare the performance of the proposed approach against the approaches cited in section 4.3. The resulting RMSE for different acceleration factors (AF) is shown in Figure 5(a) for the brain dataset and in 5(b) for the knee joint dataset.

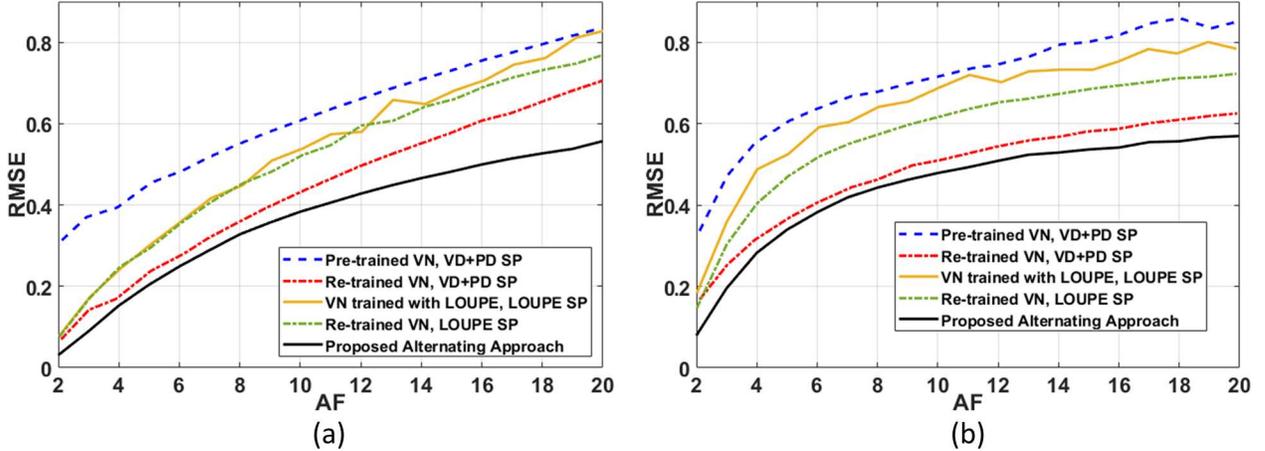

**Figure 5:** RMSE of the five compared approaches. In (a) the results for the brain dataset are shown, and in (b) the results for the knee joint dataset are shown.

The performance of the VN depends significantly on the type of SP, and also on the way it was trained. Particularly, the quality of the reconstructions was better when the VN was trained with one specific SP and that same SP was used for testing. In LOUPE, the training process randomly generates different SPs from a learned sampling density. The weakness of this approach is that the specific k-space positions of the samples are disregarded. As a result, the performance of the VN with LOUPE SP was inferior to a VN re-trained with VD+PD SP or with the proposed approach.

The proposed approach was able to surpass the performance of the other alternatives, including a VN re-trained with VD+PD SP. On average, the improvements in RMSE ranged from 14.9% to 51.2%. Note that the VD+PD SP is among the best know SPs for MRI, and it has an excellent asymptotic performance (which means it perform well with different kinds of CS reconstruction and different data). However, it is interesting that the proposed approach learned in a couple of hours a different SP that performed better than the VD+PD SP for the given data.

## 5.3. Visual comparison of the resulting images and SPs:

In Figure 6 we compared the SPs learned by the proposed approach and by LOUPE with the VD+PD SP. Note that all of them show variable density behavior, with more samples near the center of the k-space (just outside the fixed central auto-calibration area). VD+PD SP has an isometric behavior, while LOUPE and the proposed approach have not. Curiously, the proposed approach learned that more samples in the low and mid frequencies are necessary for proper reconstruction, as we will discuss this fact later. The sampling density learned by LOUPE was very similar (apart from the scaling due to AF) across different AF.

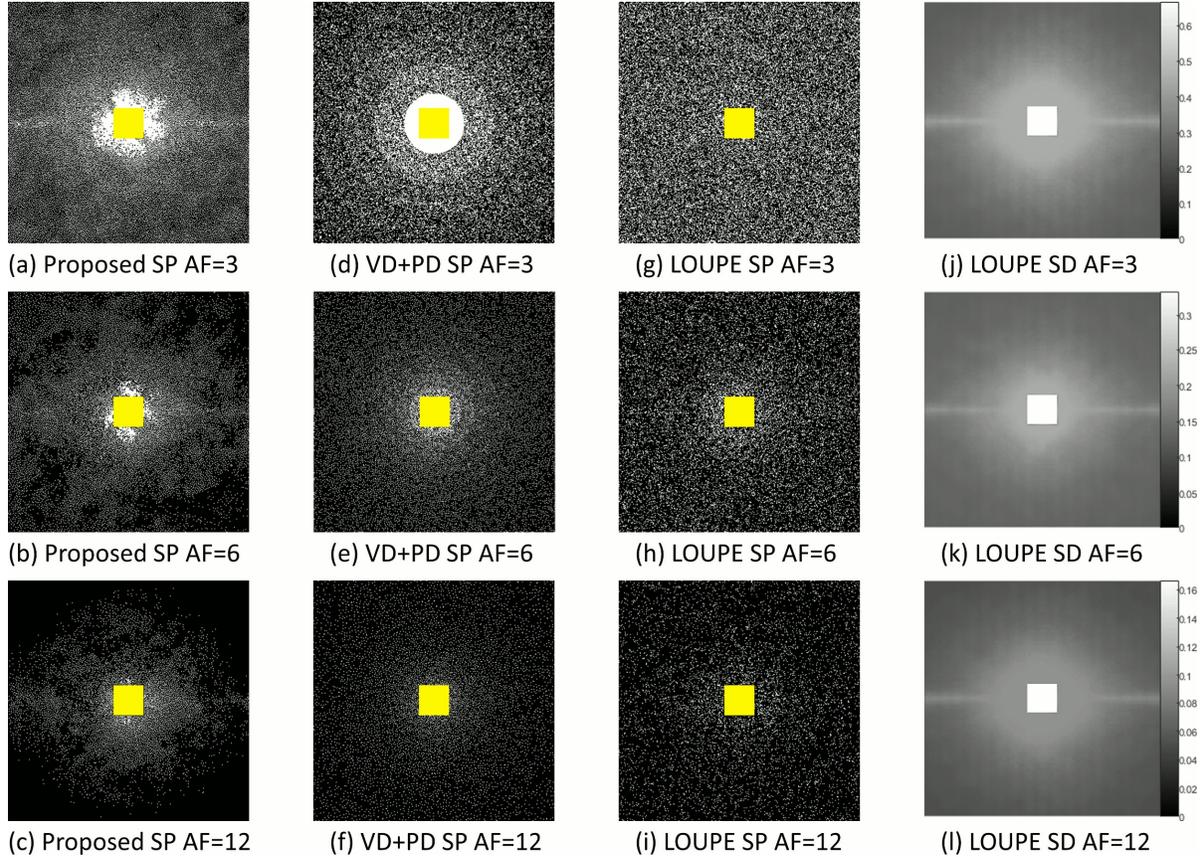

(a) Proposed SP AF=3  (d) VD+PD SP AF=3  (g) LOUPE SP AF=3  (j) LOUPE SD AF=3

(b) Proposed SP AF=6  (e) VD+PD SP AF=6  (h) LOUPE SP AF=6  (k) LOUPE SD AF=6

(c) Proposed SP AF=12  (f) VD+PD SP AF=12  (i) LOUPE SP AF=12  (l) LOUPE SD AF=12

**Figure 6:** Comparing SPs for different methods and AFs for the brain dataset. In (a)-(c) the SPs learned by the proposed approach, for AF=3, 6, and 12, are shown. In (d)-(f) the VD+PD SPs are shown, for the same AFs. In (g)-(i) the SPs generated from the sampling densities (SD) learned in LOUPE, that are shown in (j)-(l). Yellow areas correspond to fixed auto-calibration areas of the SP.

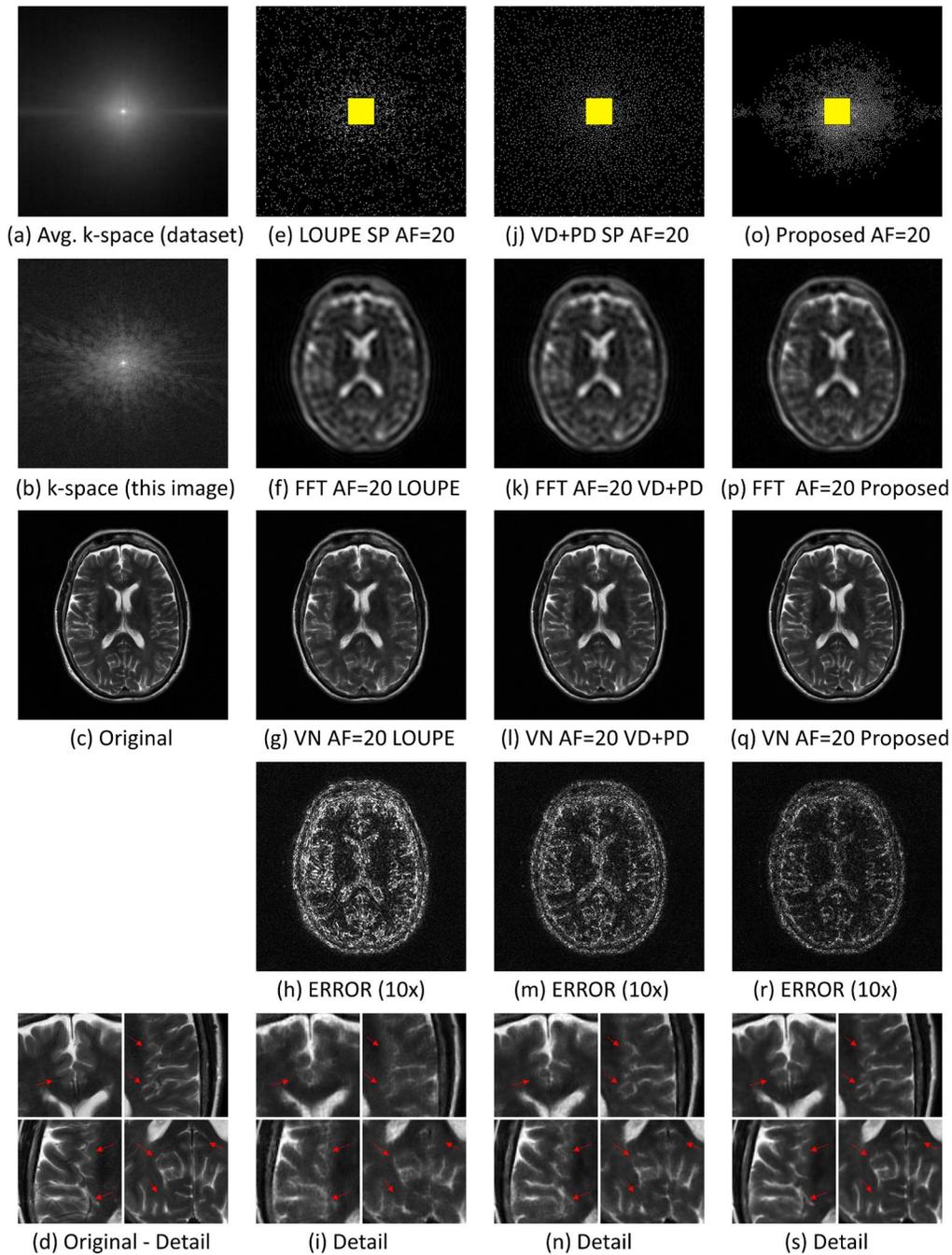

**Figure 7:** Comparing SP and images with brain dataset at AF=20. In (a) the average of the magnitude of the k-space of the training dataset is shown, while in (b) the k-space for the particular image is shown (averaged in the coil dimension). In (c)-(d) original reference image and its details are shown. In (e)-(i), the SP, FFT reconstruction, VN reconstruction, error map, and details are shown for LOUPE (RMSE=0.77). In (j)-(n), the same information for VD+PD SP are shown (RMSE=0.70), and in (o)-(s) for the proposed approach (RMSE=0.55). Arrows in the detailed images point to the most relevant differences.

In Figure 7, some visual results for the brain dataset are shown. We compared the images obtained by re-trained VN with LOUPE SP and by re-trained VN with VD+PD SP against the proposed approach. One can observe improvements obtained by the proposed approach in the reconstructed images (see details 7(d),(i),(n),(s)) and in the error maps.

One important aspect of the results in Figure 7 is regarding the FFT reconstructed images, obtained by applying $\mathbf{E}_\Omega^* \mathbf{\bar{m}}$ (see 7(f),(k),(p)). Besides being severely blurred, the FFT image obtained in the proposed approach show more recognizable structures in the middle of the blurred image. This indicates that the algorithm learned a SP that can reduce the destructive interference caused by aliasing with this kind of data. At least two visual aspects can be observed in the SPs learned by the proposed method: 1) the distribution of the samples follows the averaged magnitude of the k-space of the dataset (intensity of k-space shown in logarithmic scale in Figure 7(a)); and 2) the disposition of the samples is very specific and perhaps non-intuitive, that includes variable distance between each sample depending on the position in the k-space and the non-symmetric aspect of the SP.

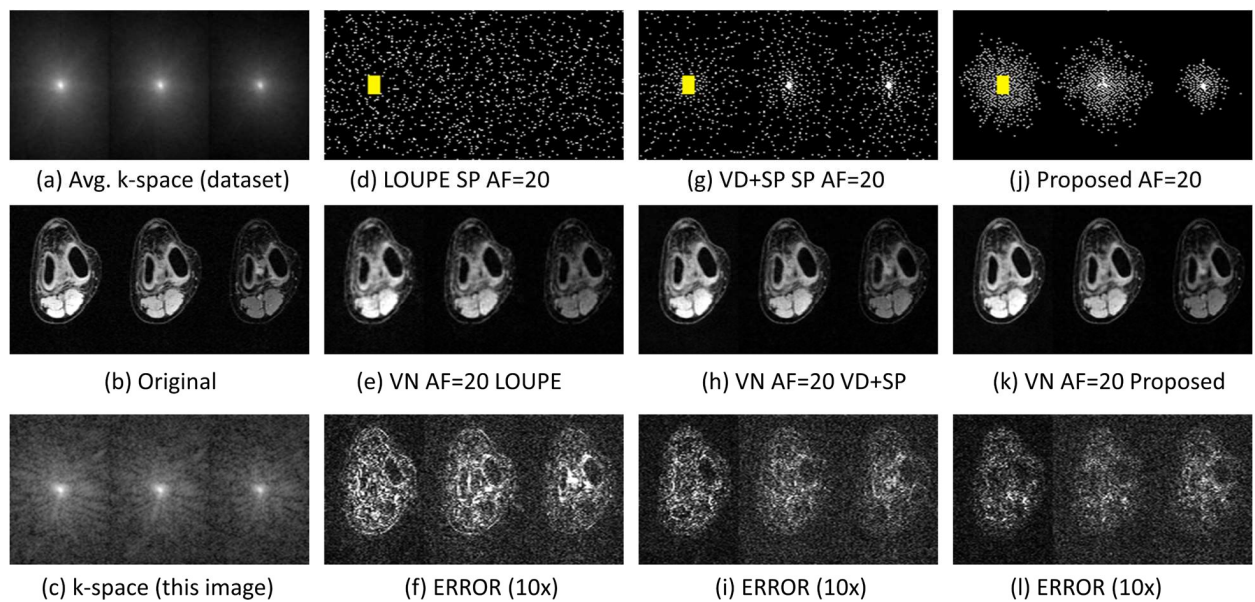

**Figure 8:** In (a) the average of the magnitude of k-space for the dataset is shown. Only three of the ten frames are shown: first, middle, last. In (b)-(c) one image and its k-space magnitude (averaged in coil dimension), used as the reference, are shown. In (d)-(f), the SP, VN reconstruction, and error map are shown for LOUPE (RMSE=0.85). In (g)-(i), the same information for VD+PD SP are shown (RMSE=0.67), and in (j)-(l) for the proposed approach (RMSE=0.62).

In Figure 8, some visual results for the knee dataset are shown, only the first, middle, and last frames are shown. Note that LOUPE learned that low-frequency central components of the other frames are not needed for reconstruction, and the VN learned how to recover them using the central area of the first frame. However, LOUPE performed worse than VD+PD SP and the proposed approach. Similar to the results with the brain dataset, here, the proposed approach also learned a SP with more samples in low and mid frequencies is necessary for reconstruction with better RMSE. However, the proposed approach also learns that a variable number of samples over time is beneficial.

## 6. DISCUSSION:

The proposed approach is an interesting alternative to jointly learn the SP and the reconstruction parameters for 3D Cartesian parallel MRI. It performed better than LOUPE, mainly because LOUPE does not learn the relative position between samples in k-space, which are relevant in parallel MRI. Some other joint approaches that can potentially be used for this problem are [44], [46], [47], however, they require non-Cartesian FFTs in their network. These approaches have more freedom in learning arbitrary k-space trajectories, while the proposed approach is limited to subset selection formulation to the SP learning, but they lose in processing speed due to the need for non-uniform Fourier transform.

The SP learning algorithm used in the proposed approach, known as BASS, has been proven effective with different CS algorithms [42] in large-scale problems. However, it is the first time it has been merged with deep learning algorithms for MR image reconstruction. In [42], it was also shown that BASS can work with different cost functions. There is no need for differentiability in BASS, which is a requirement in neural networks trained with backpropagation algorithms.

An important aspect of the joint learning approaches is the stability of the results. Due to the non-convex nature of the problem, the learning approach is likely to converge to a local minimum. In other words, several aspects of the algorithm are relevant (like the initial solution, step-sizes, decaying rates). In this sense, we observe that the proposed approach was very stable, as seen in Figure 4. The stability of the joint learning approaches is an important issue, as mentioned in [44], [47]. The forced monotonicity of the proposed joint approach is, in great part, responsible for this stability, as seen in Figure 3.

As seen in the results of figures 6, 7, and 8, the proposed approach learned SPs that are more densely sampled in low and mid frequencies than VD+PD and LOUPE SP. The learned SPs look different from the others and the disposition of the k-space samples is, to say, non-intuitive. However, the reconstructed images have better quality, showing low noise and good details. Particularly in Figure 7, the brain structures were recovered better with the proposed approach at AF=20 than with the other two pairs of VN/SP. These interesting results support the fact that joint learning of SP and neural network parameters, using algorithms like the one proposed in this paper, is important in accelerated parallel MRI.

In this work, we tested the joint learning approach with the original VN for MRI [1], however, the proposed approach is not restricted to this kind of network. Other recent deep learning networks, such as the ones discussed in [15], [16], [56], can be used.

## 7. CONCLUSION:

This paper proposes a new alternating learning approach to learn the sampling pattern and the parameters of a variational network in accelerated parallel 3D Cartesian MRI problems. The proposed approach learned a set (SP and VN) that jointly produces better images than other approaches such as LOUPE with VN or a VN with a variable density with Poisson-disc fixed SP. These results were observed with two different datasets, and performed stably across different acceleration factors, improving the quality from small AF, such as AF=2, up to very high AF, such as AF=20.

## 8. ACKNOWLEDGEMENTS

This study was supported by NIH grants, R21-AR075259-01A1, R01-AR068966, R01-AR076328-01A1, R01-AR076985-01A1, and R01-AR078308-01A1 and was performed under the rubric of the Center of Advanced Imaging Innovation and Research (CAI2R), an NIBIB Biomedical Technology Resource Center (NIH P41-EB017183). The authors are thankful for the significant contributions of Gabor T. Herman in our join previous work [42] regarding the development of the BASS algorithm. Codes for this manuscript are be available at https://cai2r.net/resources/joinly-learning-undersampling-pattern-and-variational-networks/.